\documentclass[aps,prd,superscriptaddress,nofootinbib,twocolumn,longbibliography]{revtex4-2}
\usepackage{amsmath,amssymb}
\usepackage{epsfig,graphicx}
\usepackage{xcolor}
\usepackage{natbib}
\usepackage{type1ec}

\catcode`@=11
\@addtoreset{equation}{section}
\catcode`@=12

\usepackage{hyperref}
\hypersetup{
    colorlinks,%
    citecolor=blue,%
    filecolor=blue,%
    linkcolor=blue,%
    urlcolor=blue
}
\usepackage{orcidlink}

\begin{document}

\title{\bf Interaction of Exact Gravitational Waves with Matter}
\author{Morteza Mohseni \orcidlink{0000-0002-5624-0877} }
\email{morteza.mohseni@uleth.ca,\,m-mohseni@pnu.ac.ir}
\affiliation{Theoretical Physics Group, Department of Physics and Astronomy, University of Lethbridge, 4401 University Drive, 
Lethbridge, Alberta T1K 3M4, Canada}
\affiliation{Physics Department, Payame Noor University, Tehran 19395-3697, Iran}

\author{Saurya Das \orcidlink{0000-0003-1191-8469}}
\email{saurya.das@uleth.ca}
\affiliation{Theoretical Physics Group, Department of Physics and Astronomy, University of Lethbridge, 4401 University Drive, 
Lethbridge, Alberta T1K 3M4, Canada}

%\maketitle

\begin{abstract}
    We consider interactions of exact (i.e., solutions of full nonlinear field equations) gravitational waves with matter by using 
    the Einstein-Boltzmann equation. For a gravitational wave interacting with a system of massless particles, we compute the 
    perturbed energy-momentum tensor and obtain explicit form of a set of Einstein-Boltzmann equations. We find solution to 
    this system of equations to obtain the gravitational wave profile. The interaction superposes a static term on the gravitational wave 
    profile which depends on the difference between square of the temperatures of the system in the absence and 
    in the presence of the wave. We compute this perturbed term when the states of the system obey Bose-Einstein, 
    Fermi-Dirac, and Maxwell-Boltzmann distributions, respectively. The relative strength of this term is roughly half for the 
    Fermi-Dirac, and one-third for the Maxwell-Boltzmann distributions compared with that of the Bose-Einstein distribution. We 
    consider both Minkowski and flat Friedmann-Robertson-Walker backgrounds.
\end{abstract}

%\pacs{98.80.Cq, 04.20.Fy, 04.50.Kd}
\maketitle

\section{Introduction}\label{intro}
Gravitational waves (GW) constitute an important prediction of the general theory of relativity. 
Their importance has been the main motive behind the massive century-long efforts which finally resulted in direct detection of GW 
in 2015. They provide a promising path towards new discoveries in physics and astronomy \cite{mastro_2024,2021NatRP...3..344B}. 
While there are many instances of GW in cosmology, we cite here their role in inducing the B-modes in the 
polarization of the CMB \cite{kamion}. Furthermore, the cosmological events could alter the propagation of GW, like the amplification
of the GW due to the expansion of the space-time \cite{grishchuk}.

Even though GWs are usually treated as solutions to the linearized Einsteins's field equation, exact GW,  i.e. solutions to the 
full nonlinear Einstein's field equation have also been of great interest both from physical and formal points of view. From a 
mathematical point of view, due to their unique properties, they appear in many areas of theoretical physics. Among many such 
theoretical aspects of GWs, we refer to the work of Penrose according to which around every null geodesic, any space-time can 
be approximated by a pp-wave geometry \cite{Penrose_1976}, Gibbons's work on propagation of quantized fields in non-flat 
space-times \cite{Gibbons_1975}, Zhang et.al. work on memory effect for GWs \cite{Zhang_2017}, Thorn's work on the Christodoulou 
effect \cite{Thorne_1992}, and Harte's work on strong lensing by GWs \cite{Harte_2013}. Some classic papers on GWs as exact 
solutions of Einstein's field equation and their properties are \cite{Brinkmann_1925,rosen1937plane,Bondi_1959}. A recent detailed 
review of exact GW solutions in general relativity may be found in \cite{Roche_2023}. For a pedagogical review see \cite{blau} in 
which the Penrose limit has also been discussed. For an excellent review with emphasis on non-linear nature of GWs see \cite{heis}.

A large portion of the knowledge about GWs comes from study of their interaction with matter, say, the changes in relative
distances of particles when a GW passes through them. Thus, it is interesting to ask what happens when a GW interacts with a system 
of particles. This can be broken down into two rather distinct parts: what happens to the wave, and how the system is affected. 
Assuming that the system is in an equilibrium state before the wave impinges on it, the wave moves the system out of equilibrium, and 
the internal interactions within the system bring it into a new equilibrium state. During this process, the wave may also get 
affected, namely, in the form a damping. The standard tool to study such problems is the Boltzmann equation of Kinetic theory which 
is widely used in various fields in physics, including cosmology. More specifically, within the framework of the GW investigations, 
the Boltzmann equation has been used to study interaction of GWs with matter. Damping of gravitational waves by matter was studied 
in \cite{Weinberg_2004,Watanabe_2006,Stefanek_2013,Baym_2017,Dent_2013}. The CMB temperature and polarization fluctuations due to 
inflationary gravitational waves have been studied in \cite{Atrio_Barandela_1994,Pritchard_2005}. The astrophysical Stochastic GW 
backgrounds scattering off of massive objects was studied in \cite{Pizzuti_2023}. The role of gauge invariance in the kinetic theory 
has been discussed in \cite{PhysRevE.110.025207}. In the framework of strong gravitational fields, one may mention 
\cite{Kremer_2013} where relativistic gas in a Schwarzschild space-time has been studied.

In the above references, the linearized approximation has been used to describe GWs, which is reasonable since GWs are typically 
weak. However, it is still interesting to consider a more general description. This is useful, at least from a formal point of view, 
to investigate exact GWs solutions. In fact, there are important physical aspects of GWs which are present only in full nonlinear 
theory, say, the Christodoulou effect \cite{christo}. Also there are several properties of exact GWs which, in principle, can be used in GW 
detectors like LIGO or LISA. An example of such peculiar properties is the twist of geodesic congruence discussed in \cite{shore}. Such twists
would manifest themselves in the rotation of particles in a ring of test particles due to the passage of GW which, in addition to the 
standard squeezing and squashing effect, could be used in the next generations of the GW detectors \cite{shore}. The interest in the nonlinear 
manifestations of GW is not limited to the context of the general theory of relativity \cite{lavinia}. In addition, there are in fact 
physically interesting GWs solutions that are usually treated as exact solutions of the Einstein equation. An important example of such 
solutions is the Aichulburg-Sexl (AS) space-time first introduced in \cite{Aichelburg_1971}, see also \cite{Tolish_2014} for a discussion in 
connection with the memory effect, and \cite{Mohseni_2012} for a generalization to some extended theories of gravity. The AS and its spinning 
generalization, Gyratons \cite{Frolov_2005}, are of importance in analysis of Planck scale scattering off a particle source for which the 
gravitational interactions are dominant \cite{Shore_2018}. In fact, as discussed in \cite{thooft}, the delta function singularity 
in shock wave solutions, like AS metric, postpones the non-gravitational interactions by an infinite time. This interesting property 
is not present in the linearized solutions. Another important example is the Kowalski-Glikman four-dimensional solution arising as 
a ground state of $N=2$ supergravity \cite{Kowalski_1985}. A more complicated exact solution has been presented in 
\cite{vanholten} describing the GW generated by a light wave. 

In the present work, we study the interaction of an exact GW with a system of particles using the Boltzmann equation. Our motivation for this 
study is to look for those effects of the GW-matter interaction which are only observed when the GW is treated as a solution of the 
full Einstein's field equation. In section \ref{boltzmann}, we present a brief review of the Boltzmann equation in curved 
space-times. Extensive reviews may be found in \cite{Acurdenas_2022,bruhat_7,debbasch1,debbasch2}. One of the classic papers on 
Boltzmann equation in general relativity is \cite{Lindquist_1966}. A nice discussion may also be found in \cite{Ellis_1971}. 
A conservative form of the equation suitable for numerical relativity has been presented in \cite{2014PhRvD..89h4073S}.  
In section \ref{inter} we first review the basic properties of exact GWs and then investigate the interaction with a system of 
massless particles in a static background. Relativistic jets from astrophysical objects \cite{jets} might be 
considered as an example of such systems of particles. We solve the system of Einstein-Boltzmann equations and obtain the wave profile. 
In section \ref{univ} we perform a similar analysis for a flat Friedmann-Robertson-Walker (FRW) background.
In section \ref{disc} we summarize the results.

%%%%%%%%%%%%%%%%%%%%%%%%%%%%%%%%%%%%%%%%%%%%%%%%%%%%%%%%%%%%%%%%%%%%%%%%%%%%%%%%%%%%%%%%%%%%%%%%%%%
%%%%%%%%%%%%%%%%%%%%%%%%%%%%%%%%%%%%%%%%%%%%%%%%%%%%%%%%%%%%%%%%%%%%%%%%%%%%%%%%%%%%%%%%%%%%%%%%%%%
%%%%%%%%%%%%%%%%%%%%%%%%%%%%%%%%%%%%%%%%%%%%%%%%%%%%%%%%%%%%%%%%%%%%%%%%%%%%%%%%%%%%%%%%%%%%%%%%%%%

\section{The Boltzmann equation}\label{boltzmann}

In a $3+1$ dimensional space-time $M$ equipped with a metric $g(.\,,.)$, we can describe the state of the
matter by the one-particle distribution function $f(x,p)$ which is basically the density of particles
at space-time point $x$ having a four-momentum $p$. The point $x$ belongs to $M$, the momentum $p$ resides in the 
tangent space of $M$ at $x$, and $f(x,p)$ is a non-negative function on the phase space 
${\mathcal P}_M$. Taking the constraint $p_\mu\, p^\mu=-\,m^2$ into account, the arguments of the distribution functions $f(x,p)$ belong 
to a $7$-dimensional subspace of the phase space. Regarding the momentum dependence of the one-particle distribution function, 
there are two popular choices: dependence on the contravariant components $f(x^\mu, p^i)$, and covariant components $f(x^\mu, p_i)$, 
see e.g, \cite{debbasch1,debbasch2} for a discussion on the (dis)advantages of each of these choices. Here we adopt $f(x^\mu, p^i)$.

The one-particle distribution function is usually used to construct a set of moments each with
a specific physical interpretation. In general, a moment is a totally symmetric tensor on $M$ defined by
\begin{equation}\label{mo1}
    M^{\alpha_1,\cdots,\alpha_r}=\int_{{\mathcal P}_x} f(x,p)\,p^{\alpha_1}\cdots p^{\alpha_r}\,\Omega_p
\end{equation} 
in which $p^\alpha$ are the components of momentum, and $\Omega_p$ is the volume form
$\Omega_p=\sqrt{-g}\,dp^0\,\wedge\, dp^1\,\wedge\, dp^2\,\wedge\, dp^3$. Some of the lowest order moments are, the zeroth
order, $r=0$
\begin{equation}\label{mo2}
    n(x)=\int_{{\mathcal P}_x} f(x,p)\,\Omega_p
\end{equation} 
which is the density of particles, the first moment, $r=1$, 
\begin{equation}\label{mo3}
    P^{\alpha}(x)=\int_{{\mathcal P}_x} f(x,p)\,p^{\alpha}\,\Omega_p
\end{equation} 
which is the particle four-flow, and the second moment, $r=2$
\begin{equation}\label{mo4}
    T^{\alpha\beta}(x)=\int_{{\mathcal P}_x} f(x,p)\,p^{\alpha}\, p^{\beta}\,\Omega_p
\end{equation} 
which is the energy-momentum tensor. For a system whose particles have identical mass $m$, the equation
\begin{equation}\label{eq4}
    p_\mu\, p^\mu=\,-\,m^2
\end{equation} 
confines the system states to a subset of phase space called mass shell. For mass shell, the volume form in 
Eqs. (\ref{mo1})-(\ref{mo4}) should be replaced by 
\begin{equation}\label{mo5}
    \Omega_p \rightarrow \omega_p=\frac{\sqrt{-g}}{-p_0}\,dp^1\,\wedge\,dp^2\,\wedge\,dp^3.
\end{equation} 
In the absence of non-gravitational interactions, particles move along geodesics. If we also assume that there is no collision between 
particles, e.g., when the system is a dilute gas, the evolution of the one-particle distribution function 
in terms of the time coordinate $t=x^0$ is governed by the Liouville-Vlasov equation  
\begin{equation}\label{mo6a}
    \frac{d f(x,p)}{d t}\,=\,0
\end{equation}
which is essentially stating that the number of particles is conserved. In this equation $t$ can be replaced by $\lambda$, in 
terms of which the particles trajectories are parametrized. Now, if we use the geodesic equation 
\begin{equation}\label{geod}
\frac{dp^\mu}{d\lambda}+\Gamma^\mu_{\alpha\beta}\,p^\alpha\, u^\beta\,=0,
\end{equation}
we obtain  
\begin{equation}\label{mo6}
    p^\mu\,\frac{\partial f(x,p)}{\partial x^\mu}\,-\,\Gamma^\mu_{\alpha\beta}\,p^\alpha\, p^\beta\,\frac{\partial f(x,p)}
    {\partial p^\mu}\,=\,0.
\end{equation}
When this equation holds, the moments are conserved
\begin{equation}\label{mo7}
    \nabla_{\alpha_1}\, M^{\alpha_1\,\cdots\, \alpha_r}=\,0
\end{equation}
where $\nabla$ stands for the covariant derivative with respect to space-time coordinates. The Liouville-Vlasov equation can be 
coupled with the Einstein equation by considering the energy-momentum tensor associated with $f(x,p)$ as the source in the Einstein 
equation.

When the particles undergo collisions, the system is governed by the Boltzmann equation
\begin{equation} \label{qu1}
\frac{df(x,p)}{dt}={\mathcal C},
\end{equation}
where ${\mathcal C}$ is the collision term responsible for bringing the system to an equilibrium state, see e.g., 
\cite{Acurdenas_2022}.
By using the geodesic equation governing the particles trajectories $x^\mu(\lambda)$, this can be rewritten in the following form
\begin{equation}\label{eq3}
    p^\mu\,\frac{\partial f(x,p)}{\partial x^\mu}\,-\,\Gamma^\mu_{\alpha\beta}\,p^\alpha\, p^\beta\, \frac{\partial f(x,p)}
    {\partial p^\mu}\,=\,m\,{\mathcal C}.
\end{equation}
For massless particles $m$ should be replaced by the energy.

The hydrodynamic four-velocity $u^\mu$ can be related to other physical quantities. One widely used relation is the 
\textit{Eckart's definition}
\begin{equation}\label{eq6}
    u^\mu=\frac{n^\mu}{\sqrt{-n_\mu\, n^\mu}}.
\end{equation} 
An alternative relation is the \textit{Landau-Lifshitz's definition}
\begin{equation}\label{eq8}
    u^\mu=\frac{T^{\mu\nu}\,u_\nu}{-T^{\mu\nu}\,u_\mu\, u_\nu}.
\end{equation} 
We adopt the latter definition.
%%%%%%%%%%%%%%%%%%%%%%%%%%%%%%%%%%%%%%%%%%%%%%%%%%%%%%%%%%%%%%%%%%%%%%%%%%%%%%%%%%%%%%%%%%%%%%%%%%%
%%%%%%%%%%%%%%%%%%%%%%%%%%%%%%%%%%%%%%%%%%%%%%%%%%%%%%%%%%%%%%%%%%%%%%%%%%%%%%%%%%%%%%%%%%%%%%%%%%%
%%%%%%%%%%%%%%%%%%%%%%%%%%%%%%%%%%%%%%%%%%%%%%%%%%%%%%%%%%%%%%%%%%%%%%%%%%%%%%%%%%%%%%%%%%%%%%%%%%%
%%%%%%%%%%%%%%%%%%%%%%%%%%%%%%%%%%%%%%%%%%%%%%%%%%%%%%%%%%%%%%%%%%%%%%%%%%%%%%%%%%%%%%%%%%%%%%%%%%%

\section{Interaction with gravitational waves}\label{inter}
\subsection{PP Gravitational Waves}
In the Brinkmann coordinates \cite{Brinkmann_1925}, the metric of a plane GW is given by
\begin{equation}\label{eq1}
ds^2=-K(u,x,y)\,du^2-2\,du\,dv+dx^2+dy^2
\end{equation}
in which $(u,v,x,y)=\left(\frac{t-z}{\sqrt 2},\frac{t+z}{\sqrt 2},x,y\right)$.
To be more specific, this describes a plane-fronted parallel-rays (pp) GW. Parallel-rays means that it admits a global null vector
field which is covariantly constant, and the rays are integral curves of this vector field. Plane-fronted means that wavefronts
are planar. It is interesting that the metric given in Eq. (\ref{eq1}), which is a solution to the full-nonlinear Einstein equation, still 
satisfies the linearized wave equation. An alternative coordinate system used in the literature to describe exact GWs is the Rosen 
coordinates. For an elaboration of these properties, see \cite{Roche_2023}. The relation between the Brinkmann and the Rosen 
coordinates has been discussed in \cite{blau}. This metric admits $\frac{\partial}{\partial v}$ as
a Killing vector. We have also $R^\mu_\mu=0$, and $R^\alpha_\mu R_{\alpha\nu}=0$, in which $R_{\mu\nu}\neq 0$ is the Ricci tensor. 
The metric in Eq. (\ref{eq1}) has only one non-vanishing Einstein tensor component
\begin{equation}\label{eq1gl}
    G_{uu}=\frac{1}{2}\left(\frac{\partial^2}{\partial x^2}+\frac{\partial^2}{\partial y^2}\right)\,K(u,x,y).
\end{equation}
For a GW propagating in vacuum with metric given by Eq. (\ref{eq1}), we have $R_{uu}=0$.
This is satisfied by $K(u,x,u)=K_1(u)\,(x^2-y^2)+K_2(u)\,xy$, in which $K_{1,2}(u)$ are arbitrary functions of $u$ corresponding 
to two polarizations of the wave. There is no vacuum gravitational wave solution with spherical topology in GR
while such solutions have been presented in the framework of $f(R)$ gravities \cite{ZhangH,zhanghu,liuzhang}.

In the presence of a source, we have $G_{uu}\neq 0.$ The source can be interpreted as null electromagnetic fields \cite{peres}. 
Some examples of the non-vacuum GW profiles are the AS, the four-dimensional Kowalski-Glikman wave,
and GW of a light wave. The AS metric is obtained by setting $K(u,x,y)=-\epsilon\delta(u)\ln\sqrt{x^2+y^2}$ in which $\epsilon$ is a 
constant proportional to the energy of the source particle. The Kowalski-Glikman solution is given by $K(u,x,y)=\gamma(x^2+y^2)$, 
with $\gamma$ being a constant. The GW of a light wave is the solution given in \cite{vanholten}, in which 
$K(u,x,y)=\alpha(x^2+y^2)\cos^2(k u)+F(u,x,y)$ with $\alpha, k$ being constants, and $F(u,x,y)$ satisfies the Laplace equation 
$F_{,xx}+F_{,yy}=0$. A generalized version of the AS metric has been introduced in \cite{thooft}.

%%%%%%%%%%%%%%%%%%%%%%%%%%%%%%%%%%%%%%%%%%%%%%%%%%%%%%%%%%%%%%%%%%%%%%%%%%%%%%%%%%%%%%%%%%%%%%%%%%%%%%%%%%%%%%%%%%%%%%%%%%%%%%%%%%%%
%%%%%%%%%%%%%%%%%%%%%%%%%%%%%%%%%%%%%%%%%%%%%%%%%%%%%%%%%%%%%%%%%%%%%%%%%%%%%%%%%%%%%%%%%%%%%%%%%%%%%%%%%%%%%%%%%%%%%%%%%%%%%%%%%%%%
%%%%%%%%%%%%%%%%%%%%%%%%%%%%%%%%%%%%%%%%%%%%%%%%%%%%%%%%%%%%%%%%%%%%%%%%%%%%%%%%%%%%%%%%%%%%%%%%%%%%%%%%%%%%%%%%%%%%%%%%%%%%%%%%%%%%
%%%%%%%%%%%%%%%%%%%%%%%%%%%%%%%%%%%%%%%%%%%%%%%%%%%%%%%%%%%%%%%%%%%%%%%%%%%%%%%%%%%%%%%%%%%%%%%%%%%%%%%%%%%%%%%%%%%%%%%%%%%%%%%%%%%%

\subsection{Interactions}\label{intervac}
Let us consider a system of massless particles with the distribution function $f_0(x,p)$ which is in an equilibrium 
state. Here, by massless, we mean particles of very small but not strictly vanishing masses, similar to the treatment of \cite{thooft}.
The passage of the wave changes the state of the system into a non-equilibrium one described by the distribution 
function $f(x,p)$, and finally the interactions come into play and and the system evolves into a new equilibrium 
state having the distribution function $f_e(x,p)$. The exact GW under consideration here is not asymptotically flat. Then, to give some meaning
to the passage of wave, we can assume that the wave profile is vanishing outside a specific range. This implies that the astrophysical 
sources of the this exact GW are pulses or bursts of finite duration \cite{shore}. We assume that the system initially resides in a flat 
background. In other words, we neglect the effect of the energy-momentum tensor 
\begin{equation}\label{eq18a}
    T^{(0)}_{\mu\nu}=\eta_{\mu\alpha}\,\eta_{\nu\beta}\,\int \frac{d^3p}{p^v}\,p^\alpha p^\beta f_0(x^\mu,p^i),
\end{equation} 
on the background curvature. Here, we assume the particle to move along a longitudinal axis with $p^x=p^y=0$, 
$d^3p=dp^v\,dp^x\,dp^y$, and superscript $(0)$ refers to the flat space-time. Let us, for brevity, define 
$\langle{\mathcal O}\rangle_0\equiv\int d^3p\,\frac{f_0(x,p)}{p^v}\,{\mathcal O}$ for an arbitrary function ${\mathcal O}$. Now, 
Eq. (\ref{eq4}) has $p^u=0, p^v\neq 0$, and $p^u\neq 0, p^v=0$ as solutions, corresponding to particles propagating in opposite 
directions. We choose the first one. Now, from Eq. (\ref{eq18a}), we have the following non-vanishing component of the 
energy-momentum tensor
\begin{equation}
    T^{(0)}_{uu}=\langle(p^v)^2\rangle_0\label{ttr2}.
\end{equation}
 
In the presence of GW, the energy-momentum tensor is
\begin{equation}\label{eq18}
    T^{(g)}_{\mu\nu}=g_{\mu\alpha}\,g_{\nu\beta}\,\int \frac{d^3p}{K(u,x,y)p^u+p^v}\,p^\alpha p^\beta\,f(x^\mu,p^i)
\end{equation}  
in which the superscript $(g)$ indicates the presence of GW. Now, Eq. (\ref{eq4}) gives 
\begin{equation}\label{frw44}
p^u=0,
\end{equation}
or 
\begin{equation}\label{frw45}
K(u,x,y)\,p^u+2\,p^v=0.
\end{equation} 
Choosing the first solution, we obtain
\begin{equation}
    T^{(g)}_{uu}=\langle(p^v)^2\rangle\label{ttr7},
\end{equation}
in which $\langle{\mathcal O}\rangle\equiv\int d^3p\,\frac{f(x,p)}{K(u,x,y)p^u+p^v}\,{\mathcal O}$. Thus, 
\begin{equation}\label{per4}
\delta T_{\mu\nu}\equiv T^{(g)}_{\mu\nu}-T^{(0)}_{\mu\nu} 
\end{equation}
is responsible for making changes to the GW profile. From Eqs. (\ref{ttr2}) and (\ref{ttr7}) we obtain
\begin{equation}
    \delta T_{uu}=\langle(p^v)^2\rangle-\langle(p^v)^2\rangle_0,
 \end{equation}
while all other components vanish. This is to be inserted into the following Einstein equation
\begin{equation}\label{tt4}
    G_{\mu\nu}=-\kappa\,\delta T_{\mu\nu}.
\end{equation}
Taking Eq. (\ref{eq1gl}) into account, we arrive at
\begin{equation}\label{tt4a}
    \left(\frac{\partial^2}{\partial x^2}+\frac{\partial^2}{\partial y^2}\right)K(u,x,y)=-2\,\kappa\,\{\langle(p^v)^2\rangle
    -\langle(p^v)^2\rangle_0\}.
\end{equation}

Now, we turn to the study of the evolution of the system of particles as a result of interaction with the GW. In 
the absence of the GW, we have from Eq. (\ref{mo6a})
\begin{equation}\label{eq5h}
    p^v\,\frac{\partial f_0(x,p)}{\partial v}=0.
\end{equation}
In the presence of the GW, when the system still has not reached to equilibrium, we have from Eq. (\ref{eq3}) 
\begin{equation}\label{eq15a} 
    p^v\,\frac{\partial f(x,p)}{\partial v}=\epsilon\,{\mathcal C}      
\end{equation}
Finally, when the system reaches to an equilibrium state, the collision term vanishes. Then, Eq. (\ref{eq15a}) reduces to
\begin{equation}\label{ex1}
    p^v\,\frac{\partial f_e(x,p)}{\partial v}=0.
\end{equation} 
Using a collision time approximation, one may write the collision term in Eq. (\ref{eq15a}) in the following form
\begin{equation}\label{eq14}
    {\mathcal C}\,=\,\frac{f_e(x,p)-f(x,p)}{\tau}
\end{equation}
in which $\tau$ is the collision time. We can also assume that $f(x,p)$ and $f_e(x,p)$ differ by a small amount $\varphi(x,p)$ 
such that
\begin{equation}
    f(x,p)=f_0(x,p)\{1+\varphi(x,p)\},\label{eq13}       
\end{equation}  
which is in fact the usual assumption of the Chapman-Enskog method for solving the Boltzmann equation, see e.g., \cite{Kremer_2013}.
Thus, we obtain
\begin{equation}
    p^v\,\frac{\partial\varphi(x,p)}{\partial v}\,=-\frac{\epsilon}{\tau}\,\varphi(x,p).\label{eq13w}       
\end{equation}
This equation gives $\varphi(x,p)$, which can be inserted back into Eq. (\ref{eq13}). The resulting equation leads us to
\begin{equation}
    f(x,p)=\left\{1+\exp{\left(\frac{-\epsilon\,v}{\tau p^v}\right)}\right\}f_e(x,p).\label{eq15}       
\end{equation}  

We can assume that both the initial and final equilibrium states have a Bose-Einstein distribution function, which is the case
when the system is composed of bosons. This assumption is consistent with Eqs. (\ref{eq5h}) and (\ref{ex1}). 
Thus, we have
\begin{eqnarray}
    f_0(x,p)&=&\frac{\delta(p^x)\,\delta(p^y)}{\exp{\left(\frac{\epsilon_0}{T_0}\right)}-1},\label{eq13v} \\      
    f_e(x,p)&=&\frac{\delta(p^x)\,\delta(p^y)}{\exp{\left(\frac{\epsilon}{T}\right)}-1}\label{eq16}
\end{eqnarray}
in which $T_0,T$ stand for temperatures, and $\epsilon_0,\epsilon$ represent energies. The Dirac delta functions in the right hand 
sides of the above relations guarantee the $p^x=p^y=0$ condition to hold for all particles. Thus, we obtain
\begin{eqnarray}
    \langle(p^v)^2\rangle_0&=&\int d^3p\, \frac{p^v\,\delta(p^x)\,\delta(p^y)}{\exp{\left(\frac{\epsilon_0}{T_0}\right)}-1},
    \label{eq13hh} \\      
    \langle(p^v)^2\rangle&=&\int d^3p\,\left\{ \frac{p^v\,\delta(p^x)\,\delta(p^y)}{\exp{\left(\frac{\epsilon}{T}\right)}-1}\right.
    \nonumber\\&&\left.\times \left[1+\exp{\left(\frac{-\epsilon\,v}{\tau\, p^v}\right)}\right]\right\}    \label{eq13hl}
\end{eqnarray}
We also have $\epsilon=\frac{p^u+p^v}{\sqrt 2}$, therefore
\begin{eqnarray}
    \langle(p^v)^2\rangle_0&=&\int d^3p\, \frac{p^v\,\delta(p^x)\,\delta(p^y)}{\exp{\left(\frac{p^v}{{\sqrt 2}\,T_0}\right)}-1},\label{eq131h} \\      
    \langle(p^v)^2\rangle&=&\int d^3p\,\left\{ \frac{p^v\,\delta(p^x)\,\delta(p^y)}{\exp{\left(\frac{p^v}{{\sqrt 2}\,T}\right)}-1}
    \right. \nonumber\\&&\left.\times\left[1+\exp{\left(\frac{-v}{{\sqrt 2}\,\tau}\right)}\right]\right\}.    \label{eq131l}
\end{eqnarray}
The $v$-dependence of the last term in the above equations indicate that Eqs. (\ref{eq1gl}) and (\ref{per4}) are not consistent in 
the presence of collision term. By taking the collision-less limit $\tau\rightarrow\infty$, the equations will be consistent. Thus, we obtain
\begin{equation}\label{jet1}      
    \langle(p^v)^2\rangle-\langle(p^v)^2\rangle_0=\frac{\pi^2}{3}\left(T^2-T^2_0\right)
\end{equation}
which together with Eqs. (\ref{tt4a}) gives
\begin{equation}\label{jet2}      
    K(u,x,y)=h(u)\,(x^2-y^2)+l(u)\,x\,y-K_{be}\,(x^2+y^2)
\end{equation}
in which $h(u),l(u)$ are arbitrary functions of $u$, and $K_{be}=\frac{2\kappa\pi^2}{3}\left(T^2-T^2_0\right)$. The last term on the 
right hand side can be interpreted as the perturbed part superposed on the wave as a result of interaction with the system of particles. 

If we assume that the system initial and final equilibrium states obey Fermi-Dirac distribution
\begin{eqnarray}
    f_0(x,p)&=&\frac{\delta(p^x)\,\delta(p^y)}{\exp{\left(\frac{\epsilon_0}{T_0}\right)}+1},\label{eq13ww} \\      
    f_e(x,p)&=&\frac{\delta(p^x)\,\delta(p^y)}{\exp{\left(\frac{\epsilon}{T}\right)}+1},\label{eq16ss}
\end{eqnarray}
which correspond to the case where the gas is composed of fermions like neutrinos, we can repeat the above computations to obtain
\begin{equation}\label{jet3}      
    \langle(p^v)^2\rangle-\langle(p^v)^2\rangle_0=\frac{\pi^2}{6}\left(T^2-T^2_0\right)
\end{equation}
which in turn results in the metric given in Eq. (\ref{jet2}), but with $K_{be}$ replaced by 
\begin{equation}\label{frw48}
K_{fd}=\frac{\kappa\pi^2}{3}\left(T^2-T^2_0\right).
\end{equation}
Thus, the perturbed term is weaker by a factor of $\frac{1}{2}$ compared with the Bose-Einstein case.

Finally, if we replace the above quantum distributions by their classical limit, Maxwell-Boltzmann distribution, we obtain
the same metric, this time with $K_{be}$ replaced by 
\begin{equation}\label{frw49}
K_{mb}=2\kappa\left(T^2-T^2_0\right).
\end{equation}
Here, the perturbed term is weaker compared with both quantum distributions, roughly, one-third of the BE case. 
Although the specific distributions chosen above facilitate the establishment of relations like Eq. (\ref{frw49}),
they are not exclusive assumptions for this purpose. In fact, we can repeat these calculations for other solutions of the relevant
Boltzmann equation consistent with the physical properties of the system under consideration.

The second solution to Eq. (\ref{eq4}), i.e., the dispersion relation  $K(u,x,u)\,p^u+2\,p^v=0$ does not lead to a 
set of consistent equations. 

Now, in the above equations, since the GW does not change the temperature of the gas of 
exactly massless particle by itself, the calculated perturbed part of the GW would vanish. However, if we consider the mass of 
particles to be small but not exactly zero, the above perturbed profiles will be nonvanishing. In fact, for 
particles of exactly zero mass co-propagating with the GW, there are no transverse momentum components, and the particles are
not affected by the GW as can be seen from the geodesic equations. On the other hand, by assuming a nonvanishing mass,
there will be nonzero transverse momenta which couple with the GW and there will be nonvanishing, though small, change
in the energy of the particles. For particles of finite but not very small mass, the resulting system of equations will not be consistent.
This means that massive particles can not act as a source for the exact GW under consideration.
 
Also, other factors, such as expansion of the background, can change the temperature and we can get nonvanishing results. In the 
next section, we investigate this for a flat FRW background explicitly. 

%%%%%%%%%%%%%%%%%%%%%%%%%%%%%%%%%%%%%%%%%%%%%%%%%%%%%%%%%%%%%%%%%%%%%%%%%%%%%%%%%%%%%%%%%%%%%%%%%%%
%%%%%%%%%%%%%%%%%%%%%%%%%%%%%%%%%%%%%%%%%%%%%%%%%%%%%%%%%%%%%%%%%%%%%%%%%%%%%%%%%%%%%%%%%%%%%%%%%%%
%%%%%%%%%%%%%%%%%%%%%%%%%%%%%%%%%%%%%%%%%%%%%%%%%%%%%%%%%%%%%%%%%%%%%%%%%%%%%%%%%%%%%%%%%%%%%%%%%%%

\section{Interaction in an Expanding Universe}\label{univ}
We can generalize the metric given in Eq. (\ref{eq1}) to represent an exact GW propagating in a flat FRW background.
In terms of the conformal time $d\eta=\frac{dt}{a(t)}$ in the coordinates 
$(u,v,x,y)=\left(\frac{\eta-z}{\sqrt{2}},\frac{\eta+z}{\sqrt{2}},x,y\right)$, it can be written in the following form
\begin{equation}\label{frw1}
    ds^2=a^2(-K(u,x,y)\,du^2\,-2\,du\,dv\,+\,dx^2\,+\,dy^2),
\end{equation} 
in which $a\equiv a(\eta)$ is the scale factor. Note that, to the best of our knowledge, this has not appeared in the literature, 
although, an exact anisotropic GW solution has been introduced in \cite{puetz}. The above metric does not admit $\frac{\partial}{\partial v}$
as a covariantly constant null vector. This means that the wave fronts are not planar. In other words, this does not belong to the family of 
pp-waves. It shares this property with the more general solutions presented in the latter reference.   
The nonvanishing components of the Einstein's tensor are given by
\begin{eqnarray}
G_{uu}&=&\rho+\left(K-\frac{1}{2}K^2\right)\,{\mathcal P}\nonumber\\&&
+\frac{1}{2}\,(K_{,xx}+K_{,yy})+\frac{a^\prime}{\sqrt{2}a}\,K_{,u},\label{frw2}\\
G_{uv}&=&\rho\,+\,{\mathcal P}-\frac{1}{2}\,K\,{\mathcal P},\label{frw3}\\
G_{vv}&=&\rho,\label{frw4}\\
G_{ux}&=&K_{,x}\frac{a^\prime}{\sqrt{2}a},\label{frw5}\\
G_{uy}&=&K_{,y}\frac{a^\prime}{\sqrt{2}a},\label{frw6}\\
G_{xx}&=&-\left(1-\frac{1}{2}K\right){\mathcal P},\label{frw7}\\
G_{yy}&=&-\left(1-\frac{1}{2}K\right){\mathcal P},\label{frw8}
\end{eqnarray}
in which ${a^\prime}\equiv\frac{da}{d\eta}$, ${\mathcal P}=\frac{2a^{\prime\prime}}{a}-\frac{{a^\prime}^2}{a^2}$, 
$\rho=-\frac{a^{\prime\prime}}{a}+\frac{2{a^\prime}^2}{a^2}$,  
$K_{,x}\equiv\frac{\partial K}{\partial x}, K_{,xx}\equiv\frac{\partial^2K}{\partial x^2}$, and similar expressions for $y$ 
and $u$. In these expressions, the arguments have been dropped for brevity.
The Einstein field equations for the $vv, uv, xx, yy$ and the first line of $uu$ component are satisfied by considering a perfect 
fluid with energy density proportional to $\rho$ and pressure proportional to ${\mathcal P}$. The $K$-dependent terms in these 
components are perturbations arsing from the GW. Similarly, the $ux, uy$ and the last term of the $uu$ components can be satisfied
by adding a corresponding anisotropic energy momentum tensor. The $\frac{a^\prime}{a}$ dependence of the latter energy momentum 
tensor shows that it arises due to the expansion of the background space-time. Finally, the $K_{xx}+K_{yy}$ term in the $uu$
component, which is independent of the scale factor, corresponds to the GW. If $K(u,y,y)$ satisfies the Laplace equation, it 
corresponds to propagation of sourceless GWs.  

Now, the evolution of the expanding background can be studied by specifying an equation of state ${\mathcal P}={\mathcal P}(\rho)$,
and solving the relevant perturbed Friedmann equations, which we are not interested in here. Instead, we follow the program described
in section \ref{intervac}. We consider a beam of massless particles propagating in the $z$ direction. In fact, since the metrics 
given in Eqs. (\ref{eq1}) and (\ref{frw1}) are related by a conformal transformation, null geodesics are retained. Also equation 
(\ref{frw44}) still holds. Now, from Eq. (\ref{mo6}) we obtain the following relation for the distribution function of the massless 
particles in the absence of the GW as follows
\begin{equation}\label{frw14}
\frac{\partial f_0(x,p)}{\partial v}-2\,p^v\,\frac{a_{,v}}{a}\,\frac{\partial f_0(x,p)}{\partial p^v}=0.
\end{equation}     
Similarly, in the presence of GW, we obtain
\begin{equation}\label{frw15}
    \frac{\partial f_e(x,p)}{\partial v}-2\,p^v\,\frac{a_{,v}}{a}\,\frac{\partial f_e(x,p)}{\partial p^v}=0.
\end{equation} 
The Boltzmann equations given in Eqs. (\ref{frw14}) and (\ref{frw15}) admit the following solutions
\begin{eqnarray}
[f_0(x,p)]_{be}&=&\frac{\delta(p^x)\,\delta(p^y)}{\exp{\left(\frac{a^2\,\epsilon_0}{T_0}\right)-1}}\label{frw16}\\
{[f_e(x,p)]_{be}}&=&\frac{\delta(p^x)\,\delta(p^y)}{\exp{\left(\frac{a^2\,\epsilon}{T}\right)}-1}\label{frw17}
\end{eqnarray}
respectively. Here $\epsilon_0$ ($\epsilon$) is the massless particles energy before (after) the arrival of GW, and $T_0, T$
are the corresponding temperatures. These solutions correspond to the Bose-Einstein distribution. 
There is also another set of solutions which reads
\begin{eqnarray}
    [f_0(x,p)]_{fd}&=&\frac{\delta(p^x)\,\delta(p^y)}{\exp{\left(\frac{a^2\,\epsilon_0}{T_0}\right)+1}}\label{frw18}\\
    {[f_e(x,p)]_{fd}}&=&\frac{\delta(p^x)\,\delta(p^y)}{\exp{\left(\frac{a^2\,\epsilon}{T}\right)+1}}\label{frw19}
\end{eqnarray}
and correspond to the Fermi-Dirac distribution. The third set of solutions is as follows
\begin{eqnarray}
    [f_0(x,p)]_{mb}&=&\delta(p^x)\,\delta(p^y)\,\exp{\left(-\frac{a^2\,\epsilon_0}{T_0}\right)}\label{frw20}\\
    {[f_e(x,p)]_{mb}}&=&\delta(p^x)\,\delta(p^y)\,\exp{\left(-\frac{a^2\,\epsilon}{T}\right)}\label{frw21}
\end{eqnarray}
describing the Maxwell-Boltzmann distribution. Now, we can compute the nonvanishing component of the perturbed energy-momentum 
tensor for the massless particles
\begin{equation}\label{frw22}
    \delta T^{(m)}_{uu}=\langle (p^v)^2\rangle\,-\,\langle (p^v)^2\rangle_0
\end{equation}    
in which $\langle O\rangle\equiv\int d^3p\,\frac{{\sqrt -g}f(x,p)}{a^2(K(u,x,y)p^u+p^v)}\,{\mathcal O}$,  
$\langle{\mathcal O}\rangle_0\equiv\int d^3p\,\frac{{\sqrt -g}f(x,p)}{a^2\,p^v}\,{\mathcal O}$, and the superscript 
$(m)$ refers to the massless particles. Here, ${\sqrt -g}=a^4$. By inserting Eqs. (\ref{frw16}) and
(\ref{frw17}) into Eq. (\ref{frw22}), we obtain
\begin{equation}\label{frw23}
    \delta T^{(m)}_{uu}=\frac{\pi^2}{3}\,\left(a^2\,T^2-a^2_0\,T^2_0\right)
\end{equation}    
in which $a,a_0$ are the scale factors corresponding to the temperatures $T,T_0$, respectively. 
On inserting the latter relation together with the corresponding term in the right-hand side of Eq. (\ref{frw2}), 
we obtain
\begin{eqnarray}
    K(u,x,y)&=&h(u)\,(x^2-y^2)+l(u)\,x\,y\nonumber\\&-&\frac{2\kappa\pi^2}{3}\,\left(a^2\,T^2-
    a^2_0\,T^2_0\right)\,(x^2+y^2)\label{frw24}
\end{eqnarray}
which is the same as Eq. (\ref{jet2}) modulo factors of $a^2$. Since for a system of massless particles the 
temperature goes like the inverse scale factor, the combination $a^2 T^2$ appears to be constant, and hence the perturbed part
vanishes. However, for a system of particles of small but nonzero masses, we get nonvanishing result. It is interesting that this does 
not involve derivatives of the scale factor. We can obtain equations similar to Eqs. (\ref{frw48}) and (\ref{frw49}) by using 
Eqs. (\ref{frw18}), (\ref{frw19}) and (\ref{frw20}), (\ref{frw21}), respectively. Here, we obviously have $T\neq T_0$.     

%%%%%%%%%%%%%%%%%%%%%%%%%%%%%%%%%%%%%%%%%%%%%%%%%%%%%%%%%%%%%%%%%%%%%%%%%%%%%%%%%%%%%%%%%%%%%%%%%%%
%%%%%%%%%%%%%%%%%%%%%%%%%%%%%%%%%%%%%%%%%%%%%%%%%%%%%%%%%%%%%%%%%%%%%%%%%%%%%%%%%%%%%%%%%%%%%%%%%%%
%%%%%%%%%%%%%%%%%%%%%%%%%%%%%%%%%%%%%%%%%%%%%%%%%%%%%%%%%%%%%%%%%%%%%%%%%%%%%%%%%%%%%%%%%%%%%%%%%%%

\section{discussion}\label{disc}
We obtained solutions to the Einstein-Boltzmann equations for a pp gravitational wave interacting with a collision-less system 
composed of identical massless particles propagating on Minkowski background. This investigation is well motivated based on the 
inherent nonlinear nature of the gravity, and also when the waves are not weak. The system under consideration may be used to model 
real systems like high energy astrophysical jets and gamma rays bursts. Interaction with the gravitational wave changes the state of 
the system to a new equilibrium state with different energy which in turn changes the GW profile. This causes a $u$-independent term 
to be superposed on the wave profile. The superposed term depends on $x^2+y^2$ and is proportional to the difference 
between the initial and final temperatures squared. This can be used in principle to probe variations or anisotropies in the 
background temperature. The proportionality constant is the largest when the equilibrium state
obeys the Bose-Einstein distribution, and is the smallest when it obeys the Maxwell-Boltzmann distribution. We showed that such GW 
solution can not be obtained in the presence of a collision term. We performed the same analysis for the case where the background is a 
flat FRW space-time and obtained similar results modulo overall factors of the scale factor. 
It is also interesting that, in the basic expressions appeared in the previous sections, say the Einstein or 
energy-momentum tensors and the Boltzmann equations, the GW profile $K(u,x,y)$ appeared at most in linear powers. However, the final 
results are peculiar to the exact GW solutions and can be regarded as a manifestation of the nonlinear nature of the theory.

For massive particles, this non-perturbative GW solution is not valid. It would be interesting to investigate solutions to 
Einstein-Boltzmann system in the context of extended theories of gravity to obtain GW solutions with more general topologies. It 
would also be interesting to consider GWs propagating in a non-flat expanding background.

\acknowledgments
M.M. would like to thank the Department of Physics and Astronomy, University of Lethbridge for hospitality. 
We thank the Natural Sciences and Engineering Research Council of Canada for support. We would like to thank two anonymous referees of 
Physics Letters B for invaluable comments.
\bibliographystyle{apsrev4-2}
\bibliography{boltzabrev_rev}

\end{document}